\documentclass[conference]{IEEEtran}
\usepackage{amssymb,amsmath}
\usepackage{cite}
\usepackage{graphicx}
\usepackage[caption=false]{subfig}
\usepackage{psfrag}
\usepackage{url}
\usepackage[latin1]{inputenc}
\usepackage[absolute,overlay]{textpos}
\usepackage{tikz}
\usetikzlibrary{arrows,calc,decorations.markings}
\usepackage[linesnumbered,ruled,lined]{algorithm2e}
\usetikzlibrary{topaths}
\usetikzlibrary{shapes,trees}
\usetikzlibrary{arrows}
\usetikzlibrary{shadows}
\usetikzlibrary{positioning}
\usetikzlibrary{matrix}
\usetikzlibrary{shapes.geometric}
\usetikzlibrary{decorations.pathmorphing}
\usepgflibrary{patterns}
\usetikzlibrary{calc}
\usetikzlibrary{fit}					

\usepackage{pgf}
\usetikzlibrary{arrows,automata}
\usepackage[latin1]{inputenc}

\usepackage{textcomp}
\usepackage{xcolor}
\def\BibTeX{{\rm B\kern-.05em{\sc i\kern-.025em b}\kern-.08em
    T\kern-.1667em\lower.7ex\hbox{E}\kern-.125emX}}

\usepackage{amsthm}

\usepackage{tabularx}
\usepackage{makecell}
\usepackage{multirow}

\usepackage{amsmath}
\usepackage{amssymb}
\usepackage{bbm}
\usepackage{bm}
\usepackage{comment}
\usepackage[capitalize]{cleveref}

\usepackage{pgfplots}
\usepackage{float}
\usepackage{booktabs}

\pgfplotsset{compat=1.15}

\usepackage[bottom=1.1 in,top=0.75in,left=0.63in,right=0.63 in]{geometry}

\begin{document}

\title{An Analysis of  Minimum Error Entropy Loss Functions in Wireless Communications
}
\author{
	\IEEEauthorblockN{Rumeshika Pallewela, Eslam Eldeeb and Hirley Alves \\
	}
	\IEEEauthorblockA{Centre for Wireless Communications (CWC), University of Oulu, Finland \\
	Email: firstname.lastname@oulu.fi}
}
\maketitle

\begin{abstract}
This paper introduces the minimum error entropy (MEE) criterion as an advanced information-theoretic loss function tailored for deep learning applications in wireless communications. The MEE criterion leverages higher-order statistical properties, offering robustness in noisy scenarios like Rayleigh fading and impulsive interference. In addition, we propose a less computationally complex version of the MEE function to enhance practical usability in wireless communications. The method is evaluated through simulations on two critical applications: over-the-air regression and indoor localization. Results indicate that the MEE criterion outperforms conventional loss functions, such as mean squared error (MSE) and mean absolute error (MAE), achieving significant performance improvements in terms of accuracy, over $20 \%$ gain over traditional methods, and convergence speed across various channel conditions. This work establishes MEE as a promising alternative for wireless communication tasks in deep learning models, enabling better resilience and adaptability.
\end{abstract}

\section{Introduction}\label{sec:introduction}
Over the past few decades, wireless communication systems have undergone a significant transformation. In the early stages, development was model-centric, relying on specialized knowledge of precise and well-established models. However, with the introduction of $5$G and the progression towards $6$G technologies, the emphasis has shifted towards applications that need to operate in highly complex environments. These environments are marked by unpredictable and fluctuating channel conditions, necessitating the use of more advanced and adaptive solutions.~\cite{eldeeb2022traffic}. Such drive introduces supporting new use cases, such as vehicle platooning~\cite{10028661} and unmanned aerial vehicles (UAVs)~\cite{eldeeb2023age}, with various strict QoS requirements, such as high spectral efficiency, high throughput, and low-latency in large and high dense networks~\cite{9165550}. The integration of diverse vertical services into a unified intelligent network motivates the role of deep learning (DL) in future $6$G networks~\cite{8742579}. 

DL approaches have achieved significant success in areas, such as image recognition, image reconstruction and video games which has grown severe interests in its applicability to wireless communications~\cite{8786074}. Its ability to grasp complex relationships without relying on precise mathematical models makes it particularly well-suited for addressing large-scale wireless challenges. However, applying DL in wireless communication applications imposes several challenges, such as complexity, explainability, flexibility and performance stability, making it vastly used in wireless communications-related tasks.

\begin{figure}[t]
    \centering
    \includegraphics[width=1\linewidth]{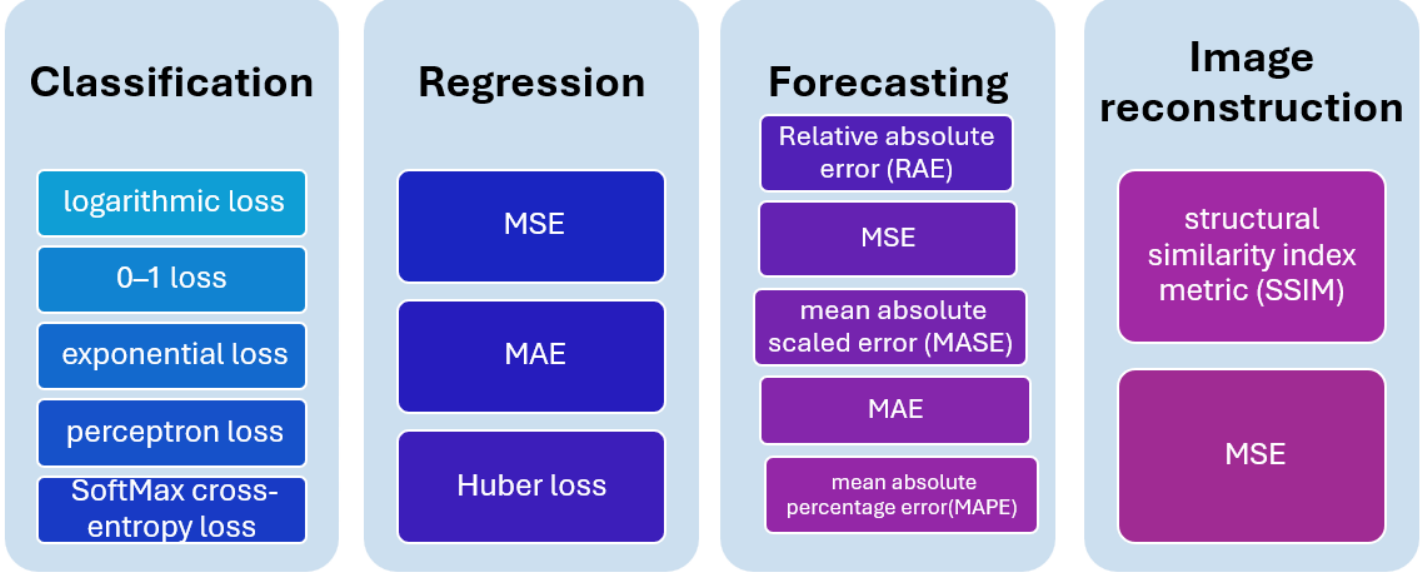}
    \vspace{-2mm}
    \caption{Different loss functions used in DL}
    \vspace{-2mm}
    \label{losses}
\end{figure}

Among these challenges is designing a proper information-theoretic error criterion (loss function) that can adapt well to the demands of wireless communication tasks. As shown in Fig.~\ref{losses}, an enormous amount of loss functions have been introduced to several DL problems. However, none were designed, specifically, to serve the demands of wireless communication~\cite{7765094}. To mention a few, semantic and task-oriented  communication~\cite{eldeeb2024multitaskorientedsemanticcommunication}, which suggests transmitting a portion of the data needed for a specific task instead of the whole data, still faces difficulties in designing a proper error criterion instead of traditional DL loss functions. Other applications, such as channel estimation and DL-based transceivers, rely on existing loss functions instead of designing a proper criterion that serves its demands.

To this end, minimum error entropy (MEE) stands out as an information-theoretic criterion that effectively captures the intricate, higher-order information embedded within the error distributions~\cite{Hu_Fan_Wu_Zhou_1970}. By incorporating the higher-order statistical features of the error signal, MEE criteria are particularly effective at handling outliers, making them highly efficient in managing impulsive noise~\cite{1011217}. This work aims to analyse an MEE criterion with reduced computational complexity for various wireless communication applications.



\subsection{Literature Review}
\vspace{-0mm}

As shown in Fig.~\ref{losses}, mean absolute error (MAE) and mean squared error (MSE) are widely regarded as robust indicators of prediction accuracy, particularly in regression tasks. MAE, which reflects the concept of average error, is frequently used to assess the performance of multivariate regression models. In contrast, MSE has been recognized as the key quantitative metric in signal processing, serving as the benchmark standard for signal quality and precision measure.

However, as the requirements of DL applications started to become very specific~\cite{Erdogan_Yoshioka_2018}, MSE and MAE deemed too generalized. Therefore, several loss functions have been recently proposed for different applications. In~\cite{9761227}, the authors proposes an adaptive loss function for channel estimation. The work in~\cite{CHEN2022121808} designs a loss function to be used in wind prediction applications specifically. However, these loss functions do not capture the high order statistics of the data, posing challenges performing in noisy and outliers scenarios. The authors in~\cite{Santos2005TheEE}, introduces MEE, a higher order loss function, primarily for time series forecasting. In addition, MEE has also been utilized in classification problems~\cite{9445619} and image compression~\cite{4107016}. In this work, we adopt the MEE criterion for various wireless communication applications.


\vspace{0 mm}
\subsection{Contributions}
The major contributions of this paper are summarized as follows:
\begin{itemize}
    \item We evaluate the performance of the MEE loss function against traditional loss functions, such as MAE and MSE, focusing on accuracy and robustness.
    
    \item We examine the theoretical aspects of MEE, taking into account its time complexity and accuracy.

    \item We showcase the performance of the MEE loss function on two wireless communication applications; over-the-air regression and indoor localization.

    \item Simulation results show that the MEE loss function outperforms existing functions in different channel realizations. The MEE model achieves at least $20 \%$ gain over existing loss functions.

\end{itemize}
To our knowledge, this is the first work to adopt the matrix based MEE criterion for deep-learning applications related to wireless communications.

\subsection{Outline}
Section~\ref{sec:sysmodel} shows the system model. The proposed MEE algorithm is presented in Section~\ref{sec:MEE}. Section~\ref{sec:results} evaluates the proposed model through simulation analysis. Finally, Section~\ref{sec:conclusions} concludes the paper and discusses potential future works.

\vspace{0 mm}
\section{System model and Problem Formulation}\label{sec:sysmodel}
\begin{figure}[t]
\centering
\includegraphics[width=1\columnwidth,trim={0 6.5cm 0 5cm},clip]{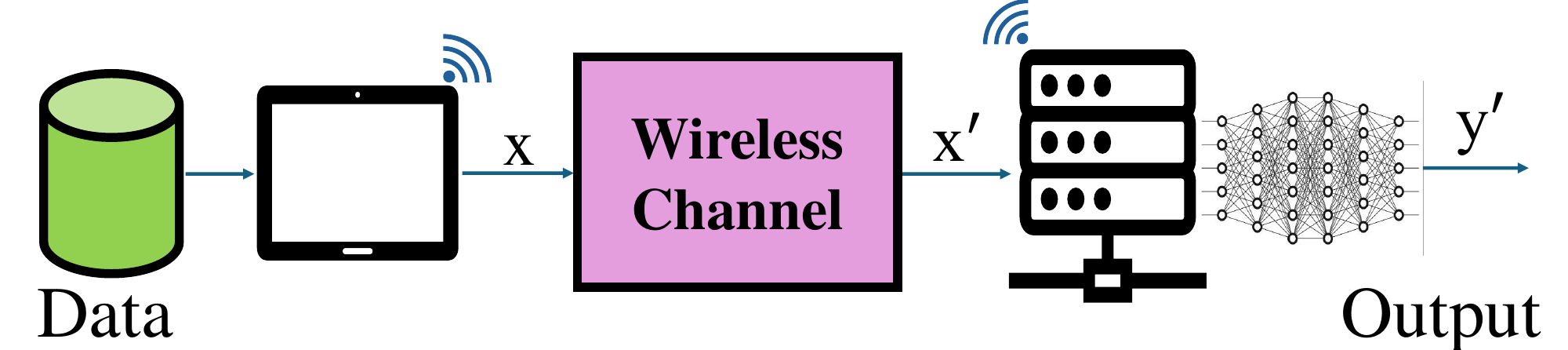}
\centering
\vspace{-2mm}
\caption{The proposed system model involves a user transmitting a signal through a wireless channel to a CPU, which then predicts the corresponding target output.}
\vspace{-2mm}
\label{chap2}
\end{figure}

\vspace{0 mm}
Consider the wireless communication system presented in Fig.~\ref{chap2}, emphasizing the data flow from an end device to a central processing unit (CPU) via the wireless channel. Then, let $\mathrm{x}$ be the transmitted signal from the device and $\mathrm{x}^{\prime}$ denotes the received signal at the CPU. The wireless channel between the transmitter and the receiver is modeled as
\begin{equation}
    \label{AWGN_Channel}
    \mathrm{x}^{\prime} = \mathrm{h} \: \mathrm{x} + \mathrm{n},
\end{equation}
where $\mathrm{h}$ is the channel that can be modeled as Rayleigh flat fading and $\mathrm{n} \sim \mathcal{N}(0,\sigma^2)$ is additive white Gaussian noise (AWGN) with power $\sigma^2$.

To this end, the CPU uses the received signal $\mathrm{x}^{\prime}$ to perform a specific task, using some non-linear function $g$ (\emph{e.g.}, a neural network), and estimate an output $\mathrm{y}^\prime = g(\mathrm{x})$, corresponding to a true output $\mathrm{y}$. 
Let the the non-linear function $g$ be a neural network (NN), then
\begin{equation}
\mathrm{y}^{\prime} = \mathrm{x}^{\prime} \: \mathrm{w} + \mathrm{b},
\end{equation}
where $\mathrm{w}$ represents the weights of the neural network, and $\mathrm{b}$ indicates the bias term of the neural network.
Given the inherent challenges and limitations in wireless communication, especially the imperative to minimize the gap between true and predicted targets, it is vital to devise sophisticated techniques for processing and analyzing the data received at the receiver end. Therefore, accurately interpreting the received signal $\mathrm{x}^{\prime}$ is essential for robust system performance.

\subsection{Problem Formulation}
 
The weights $\mathrm{w}$, which the neural network model learns, are crucial parameters that characterize how well the model predicts the outputs. Neural networks rely on updating the model weights by minimizing a simple error measurement $\mathcal{L}(\mathrm{y}^{\prime}, \mathrm{y})$, which can be represented as the deviation between the true output $\mathrm{y}$ and the predicted output $\mathrm{y}^{\prime}$. This update is carried out using the gradient of the loss function and can be expressed through a basic step of stochastic gradient descent (SGD)
\begin{equation}
\mathrm{w} \leftarrow \mathrm{w} - \eta\nabla \mathcal{L}(\mathrm{y}^{\prime}, \mathrm{y}),
\end{equation}
where $\eta $ represents the learning rate and ${\nabla}\mathcal{L}(\mathrm{y}^{\prime},\mathrm{y})$ indicates the gradient of the chosen loss function. 

Our goal is to identify a precise information-theoretic loss function for updating the neural network's weights, thereby minimizing trivial errors (\emph{i.e.}, $|\mathrm{y} - \mathrm{y}^{\prime}|$). The optimization problem is formulated as
\begin{equation}
\mathbf{P1:}\qquad \underset{ \mathcal{L}(\mathrm{y}^{\prime}, \mathrm{y})}{\min}       \ \ \  \sum_{n=1}^N |\mathrm{y}_n - \mathrm{y}^{\prime}_{n}|,
\end{equation}
where $N$ denotes the size of the training data. This optimization problem targets finding a loss function $\mathcal{L}({y}^{\prime}, {y})$ that can minimize the L$1$ deviation of the target output and the estimated output~\cite{Pallewela2024}.

\section{The MEE Criterion}\label{sec:MEE}
Consider Renyi's entropy $\mathrm{H(e)}$,
which can be computed using Gaussian kernels through Parzen windowing~\cite{1011217}. This approach results in the quadratic form of Renyi's entropy and involves kernel density estimation (KDE)
\begin{equation}
\label{eq:generaleq}
H_2(e) = -\log \int_{-\infty}^{\infty} {P_{e}(e)}^2 \, de,
\end{equation}
where $P_{e}(e)$ represents the probability density function (PDF) of the error and the term $\log \int_{-\infty}^{\infty} {P_{e}(e)}^2 \, de$ corresponds to the information potential (IP). The PDF can then be estimated using the Parzen density estimator, with the assumption that the kernel function is non-negative. Therefore, the MEE criterion can be rewritten as 
\begin{equation}
\hat{H}_2(e) = -\log \left(\frac{1}{N^2}\right) \left(\sum_{i=1}^{N} \sum_{j=1}^{N} k_\sigma(e_i - e_j)\right),
\label{second_order_rey_entr}
\end{equation}
where $e_i$, $e_j$ are the sampled components of the error signal and $k_\sigma(.)$  denotes the window function with a bandwidth $\sigma$. The most common window function used in MEE is the Gaussian function $G_\sigma (e) = \exp \left(\frac{-e^2}{2\sigma^2}\right)$~\cite{8264748}. The target in neural networks is to find the set of optimal weights $\mathrm{w}^*$ that can minimize a specific loss
\begin{equation}
    \mathrm{w}^* = \arg\min_{\mathrm{w}} H(e).
\end{equation}

Reyni's entropy is used to obtain the general MEE loss function. However, it suffers from high complexity in higher-order dimensions~\cite{6954500}. The computational cost of the current form of MEE is driven by the double summation over the error signal samples, resulting in $N^2$ operations, where $N$ is the number of samples. This includes calculating the Gaussian window function $G_\sigma(e_i - e_j) = \exp \left( -\frac{(e_i - e_j)^2}{2\sigma^2} \right)$ for each sample pair of $i$ and $j$. Consequently, the overall complexity of the function is $O(N^2)$, making MEE computationally intensive for large datasets. Therefore, we present a less complex version of the MEE criterion to mitigate this situation.
In~\eqref{second_order_rey_entr}, a reproducing kernel Hilbert space (RKHS) can be applied to the data as a mapping function 
\begin{equation}
\phi: X \rightarrow H, \: \text{with} \: \: \kappa(x, y) = \langle \phi(x), \phi(y) \rangle,
\end{equation}
where $H$ is the RKHS, $\kappa(x, y) = \int_X h(x, z)h(y, z)dz$, and $h(x, y)$ is a windowing function. 
By forming the Gram matrix $K$ with elements $K_{ij} = \kappa_{2\sigma}(x_i, x_j)$, Reyni's entropy (utilizing  Parzen windowing) is formulated in a matrix-based form
\begin{equation}
\hat{H}_2(x) = - \log \left( \frac{1}{n^2} tr(KK) \right) + C(\sigma),
\end{equation}
with $C(\sigma)$ considers the normalization factor, ensuring the Parzen window integrates into one. A relationship between the IP estimator and the Frobenius norm of the Gram matrix ${K}$ can be formulated using $\|K\|^2 = tr(KK)$. Then,
\begin{equation}
\sum_{i=1}^{N} \lambda(K)^2 = tr(KK),
\end{equation}
where the normalization of a positive definite matrix can be represented as~\cite{6954500}
\begin{equation}
\hat{A}_{ij}^{(\ell)} = \frac{A_{ij}^{(\ell)}}{\sqrt{A_{ii}^{(\ell)} A_{jj}^{(\ell)}}},
\end{equation}
where $A_{ij}^{(\ell)} = \exp\left( -d_{\ell}^2(\mathrm{x_i}, \mathrm{x_j}) \right)$ is a matrix and $\hat{A}^{(\ell)}$ is the normalized version.

Therefore, the MEE loss can be formulated using a matrix form as
\begin{equation}
\mathcal{L}_{\text{MEE}}( \mathrm{y}^{\prime}, \mathrm{y}) = \frac{1}{2} \log_2 \left( \sum_{i=1}^{N} \lambda_i(A)^2 \right),
\label{eq:LMEE}
\end{equation}
where $A$ is a normalized positive definite matrix  calculated as $A_{ij} = \frac{1}{N} \frac{K_{ij}}{\sqrt{K_{ii} K_{jj}}}$, and $\lambda_i(A)$ is the $i$-th eigenvalue of $A$. Here, $K$ is the Gram matrix obtained by evaluating a positive-definite kernel $\kappa$ with $e_i = \mathrm{y}_i - g(\mathrm{x}_i)$ is the residue of the model for each pair of the training data points. 
This formulation decreases the complexity of the MEE loss in comparison to the original Kernel-based MEE. In addition, it facilitates using the MEE criterion in handling high dimensional problems. Finally, we iteratively update the kernel bandwidth $\sigma$ after each training iteration
\begin{equation}
\sigma = \text{median}\left(\{(e_i - e_j)^2\}_{i,j=1}^{N}\right),
\label{adaptivekernel}
\end{equation}
which stabilizes the MEE performance~\cite{silvestrin2023revisiting}. \textbf{Algorithm~\ref{alg1}} illustrates the matrix-based MEE algorithm.

\begin{algorithm}[!t]
\SetAlgoLined
\textbf{Input:} Training dataset, number of training epochs $E$ and number of gradient steps $G$

\textbf{Initialize:} neural network weights ($\mathrm{w}$) and kernel bandwidth $\sigma$

\For{\text{epochs} $e$ in $\{1$,...,$E$\}}{
\For{\text{gradient step} $g$ in $\{1$,...,$G$\}}{

Sample a batch from the dataset

Calculate the neural network output by performing a feedforward pass


Calculate the error of the NN $e=|\mathrm{y} - \mathrm{y}^{\prime}|$

Compute the pairwise distance of the errors $\|e_i - e_j\|^2$

Compute the MEE Loss using~\eqref{eq:LMEE}

Optimize and update the weights using Adam optimizer

Update $\sigma$ estimate using~\eqref{adaptivekernel}

}
}

\textbf{Return} Optimized NN weights

\caption{The matrix-based MEE algorithm.}
\label{alg1}  \vspace{-1mm}
\end{algorithm}
 \vspace{-1mm}

\begin{figure*}[t]
\centering
\subfloat[Ideal\label{fig:ideal_channel}]{\includegraphics[width=0.335\textwidth,trim={0 0 0 0},clip]{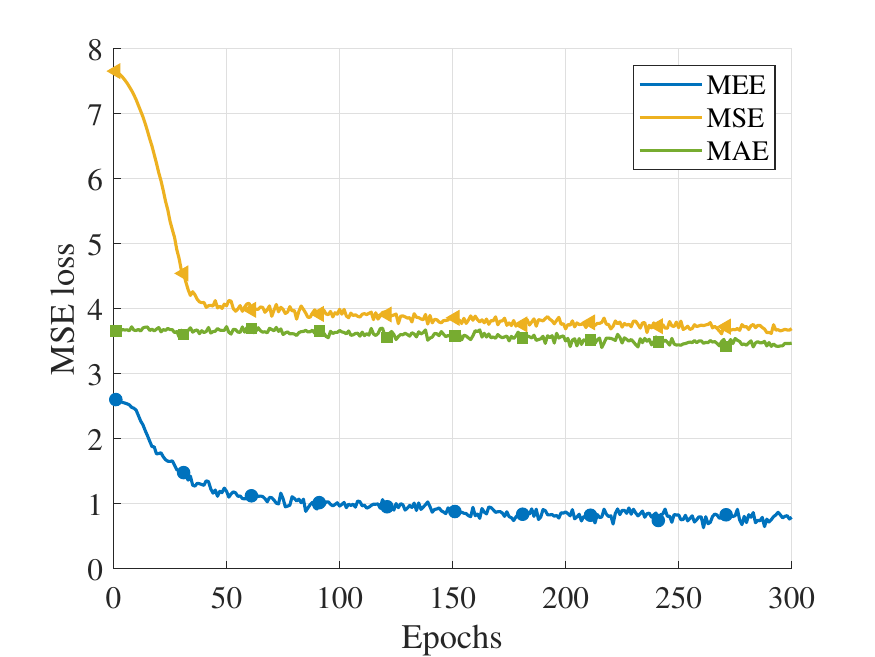}}
\hskip -2.28ex
\subfloat[AWGN\label{fig:awgn_channel}]{\includegraphics[width=0.335\textwidth,trim={0 0 0 0},clip]{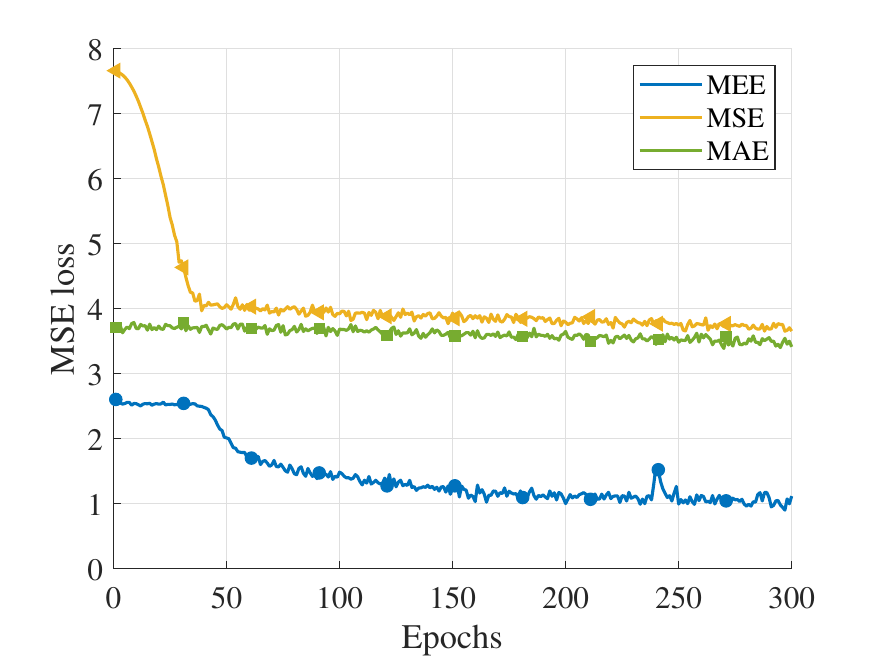}}
\hskip -2.28ex
\subfloat[Reyleigh fading\label{fig:rayleigh_channel}]{\includegraphics[width=0.335\textwidth,trim={0 0 0 0},clip]{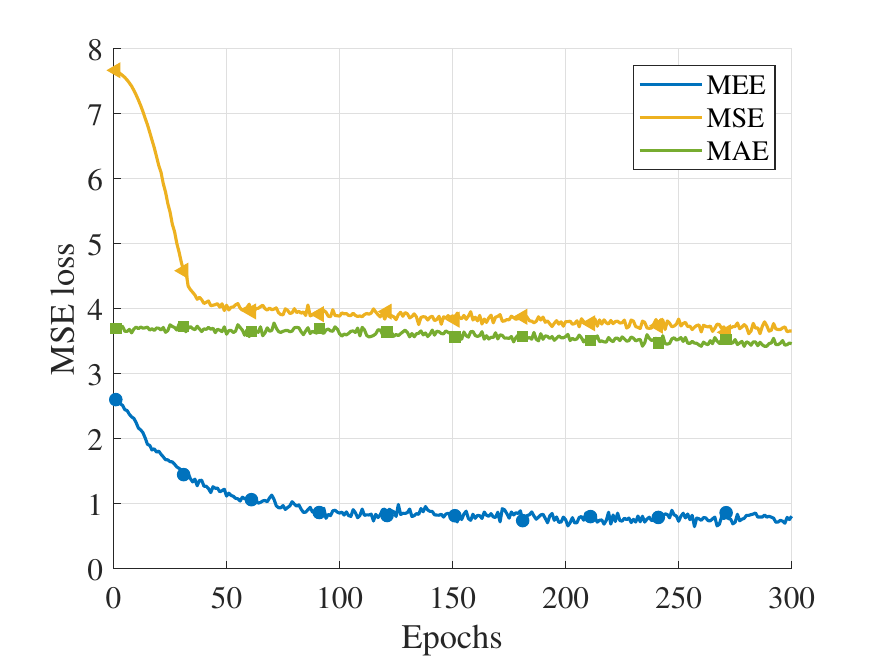}}
\caption{The comparison of MEE, MSE, and MAE setting SNR to $5$ dB using three different channel conditions: (a) ideal channel, (b) AWGN channel, and (c) Rayleigh fading channel.}
\label{fig:comparison}
\end{figure*}

\section{Selected Use Cases and Discussion}\label{sec:results}

This section evaluates the proposed MEE loss function compared to existing loss functions on two wireless communication use cases: \textit{i)} over-the-air regression, and \textit{ii)} indoor localization. We carried out our simulation using the Puhti supercomputer on the CSC server, leveraging instances configured with $32$ processing cores, $60$ gigabytes (GB) of RAM, and $60$ GB of local disk space. This high-performance setup provided the necessary computational power and storage capacity to efficiently handle the demands of our simulation tasks.

\subsection{Case Study: Regression}\label{sec:MEE_2}
In this case study, we investigate the proposed MEE criterion on over-the-air regression. Consider the communication system presented in Section~\ref{sec:sysmodel}, where we assume that the CPU performs a regression prediction task on the received message and transmits it back to the end device. We use the ``\textbf{jh-simple}'' dataset~\cite{Heaton}, which consists of columns providing information about income, aspect, save rate, subscriptions job, and area of individuals. The CPU performs a regression task to estimate the age of each person. We use $256$ QAM modulation to simulate the transmission of the dataset features from the end device to the CPU. To perform the regression task, the CPU utilizes a neural network of $4$ layers with sizes ($20,20,30,1$), respectively. Table~\ref{tab:parameters} summarizes the neural network parameter used for this use case. We compare the proposed MEE loss function to the baseline loss metrics during training.

\begin{table}[t]
\centering
\caption{The model parameters used in the simulation.}
\label{tab:parameters}
\begin{tabular}{|c|c|c|}
\hline
\textbf{Parameter} & \textbf{Use case 1} & \textbf{Use case 2} \\
\hline
Number of epochs   & $300$          & $500$\\
\hline
Batch size         & $32$           & $32$ \\
\hline
Optimizer          & Adam           & Adam\\
\hline
Learning rate      & $0.0005$       & $0.00005$ \\
\hline
\end{tabular}
\end{table}


Fig.~\ref{fig:comparison} compares the proposed MEE criterion to MSE and MAE in three cases: \textit{i) perfect channel}, where we assume perfect features transmission; \textit{ii)} AWGN channel, where we set the target SNR to $5$ dB; and \textit{iii) Rayleigh fading}, where we add a Rayleigh flat fading to the AWGN channel with SNR of $5$ dB. Across all scenarios presented, the MEE consistently outperforms MSE and MAE in terms of achieving the lowest loss values. MEE demonstrates rapid convergence within approximately $50$ epochs, showcasing its robust capability to effectively handle various channel impairments.

Conversely, the MAE shows poor convergence, which suggests that MAE is less sensitive to changes in error magnitudes, which might limit its adaptability in environments with complex or noisy dynamics. Meanwhile, MSE shows a steeper initial decrease in loss than MAE, but tends to stabilize at higher values than MEE. This experiment indicates that in low SNR scenarios, \emph{e.g.,} $5$ dB, the proposed MEE criterion outperforms MSE and MAE.

\subsection{Case Study: Indoor Localization}\label{sec:MEE2}
In this case study, we focus on indoor localization to evaluate the proposed MEE loss function. The global positioning system (GPS) ensures high localization accuracy in outdoor settings. It is indispensable for navigation, tracking, and positioning applications, but its performance deteriorates indoors due to signal blockages and weakening~\cite{8662548}. Precise positioning is especially vital in these challenging environments, and implementing robust localization algorithms and models to minimize errors is increasingly essential.



Consider the communication system presented in Section~\ref{sec:sysmodel}, where the CPU targets to determine the position of indoor users based on their received signal strength indicator (RSSI). Consider an environment with $L$ access points (APs), where each AP $l$ has a location $\{u_l, v_l\}$, and $U$ users, where each user $u$ has a location $\{u_u, v_u\}$. The CPU records the RSSI measurements, denoted as $\mathrm{x}$ of each user with respect to each AP. In this use case, the RSSI measurements are tte transmitted features, which are the input of the neural network, while the user location is the output of the neural network. This method simplifies the localization process, enabling a straightforward determination of the target's exact position using RSSI measurements and a regression model. The CPU  maintains client connections in both uplink and downlink. In downlink communication, data is broadcast by the CPU to all the clients, whereas in uplink communication, all the clients utilize one Gaussian multiple-access channel.


In this use case, we use the dataset ``\textbf{UJIIndoorLoc}''~\cite{torres14}, which consists of $520$ APs and $20$ users distributed over $108703$ squared meter area. First, we perform some preprocessing steps to identify APs that are not in range during measurements, where we drop all the APs that do not have any RSSI data or have a drop rate of $98 \%$. The initial number of APs is reduced from $520$ to $248$.

Table~\ref{tab:parameters} summarizes the hyperparameters utilized in the indoor localization user case. The neural network utilized in the simulation comprises $5$ FC layers ($256$,$128$,$64$,$2$). We use the ReLU activation layer after each FC layer. A softmax layer is used after the output layer. We evaluate the model using the mean error distance (MED), which is computed by averaging the euclidean distances between the reference positions and the predicted positions for each sample~\cite{9708854}
\begin{equation}
\text{MED} = \frac{1}{N} \sum_{i=1}^{N} \text{dist}(\mathrm{y}_i, \mathrm{y}^{\prime}_i).
\end{equation}


\begin{figure}[t]
\centering
\includegraphics[width=1\columnwidth]{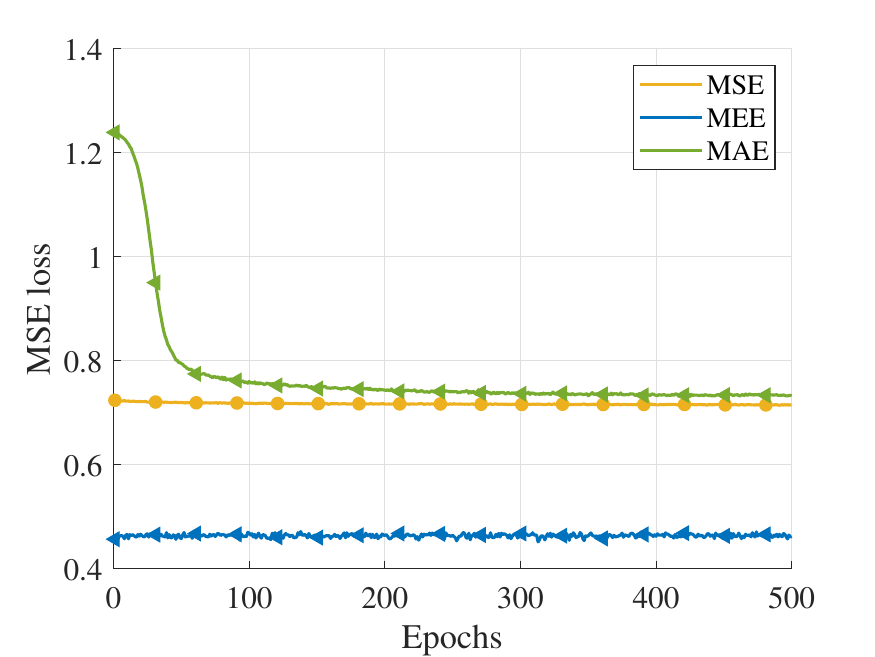}
\centering
\vspace{-2mm}
    \caption{Convergence of the MEE loss function compared to MAE and MSE loss functions during training for the localization use case as a function of the training epochs.}

\vspace{-2mm}
\label{localization1}
\end{figure}

Fig.~\ref{localization1} reports the loss values for the proposed MEE loss function compared to MSE and MAE obtained during training for $500$ epochs.
Initially, all three loss functions start at higher values, with MAE beginning at approximately $1.2$, followed by MSE and MEE. The stability of the MEE curve, which is the flattest after convergence, indicates a minimal and consistent loss. In contrast, MSE and MAE exhibit minor fluctuations but remain generally stable after their initial convergence. This figure underscores the effectiveness of MEE in minimizing loss, demonstrating superior performance compared to MSE and MAE in this training phase.
In addition, the MEE converged to lower loss values compared to MSE and MAE.

\begin{figure}[t]
\centering
\includegraphics[width=1\columnwidth]{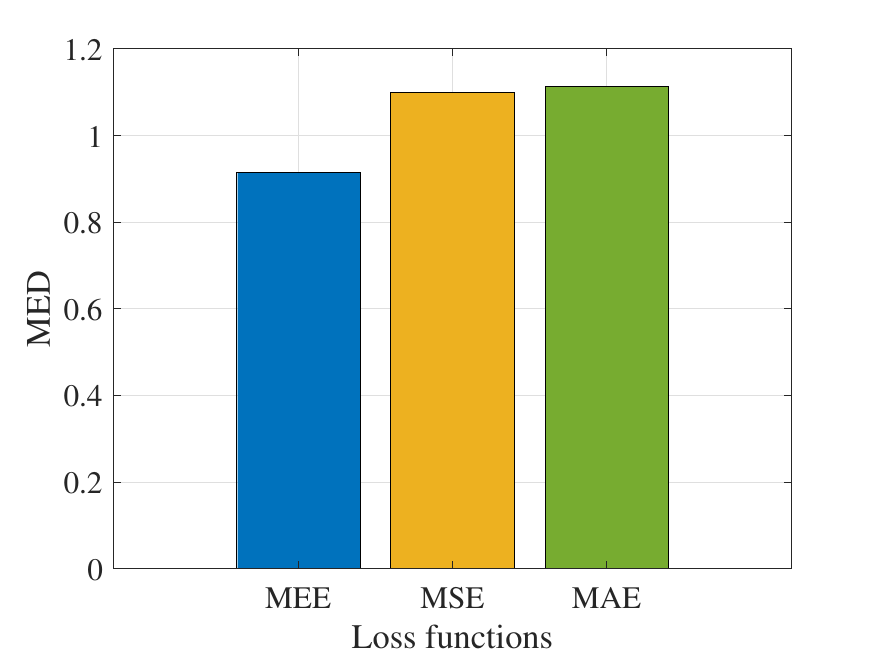}
\centering
\vspace{-2mm}
    \caption{Comparison of the MED for the proposed MEE loss function compared to MSE and MAE loss functions. three different loss functions-MEE, MSE and MAE-in the context of the localization use case.}

\vspace{1mm}
\label{local2}
\end{figure}
Fig.~\ref{local2} evaluates the MED of the proposed MEE criterion compared to MSE and MAE over a set of unseen test data points. We notice that MEE provides the best MED results ($0.9$) on new unseen data, reflecting better generalization than MSE and MAE, which both record around $1.1$ MED values.

\begin{figure}[t]
\centering
\includegraphics[width=1\columnwidth]{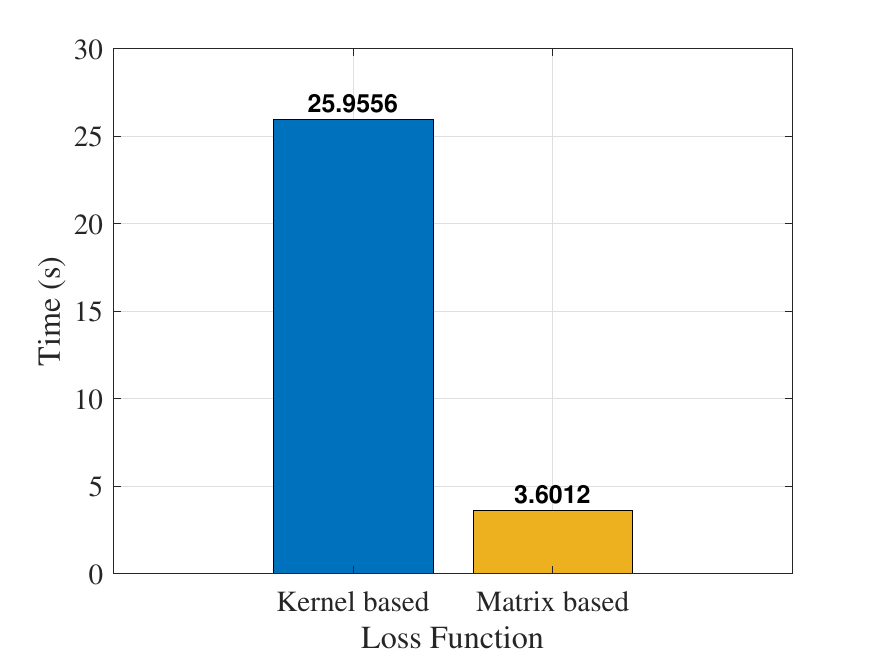}
\centering
\vspace{-2mm}
\caption{Time complexity analysis for kernel-based and matrix-based MEE loss function to train $10$ epochs.}
\vspace{-2mm}
\label{fig:complextime}
\end{figure}

\subsection{Complexity Analysis}
In this subsection, we compare between the time complexity of the traditional kernel-based MEE and the proposed matrix-based MEE. We measure the required time to train $10$ epochs of each MEE method. As shown in Fig.~\ref{fig:complextime}, we can notice that the matrix-based method only needs $3.60$ s to complete $10$ epochs of training compared to the kernel-based method, which consumes $25.96$ s.


\section{Conclusions}\label{sec:conclusions} 

This paper presents a MEE loss function designed specifically for regression analysis applications in deep learning for wireless communications. Based on Renyi's entropy, the proposed MEE criterion outperforms traditional loss functions, such as MSE and MAE. The proposed model is very effective in noisy scenarios and in managing outliers. Furthermore, we introduce a less complex version of the MEE criterion for practical scenarios. The MEE criterion demonstrates enhanced adaptability, accuracy, and resilience across various wireless communication scenarios through regression and indoor localization simulations. The proposed model enhanced the prediction accuracy by at least $20 \%$ over baseline schemes. MEE can also be adapted for other wireless communication domains, such as image reconstruction problems, reinforcement learning, and distributed learning, which are left for future work.


\appendices 

\section*{Acknowledgments} \vspace{1mm}
This research was supported by the Research Council of Finland (former Academy of Finland) 6G Flagship Programme (Grant Number: 346208).

\bibliographystyle{IEEEtran}
\bibliography{IEEEabrv,references}
\end{document}